%% file: dantzig_article.tex
\documentclass[11pt]{article}

\usepackage{PRIMEarxiv}

\usepackage[utf8]{inputenc} % allow utf-8 input
\usepackage[T1]{fontenc}    % use 8-bit T1 fonts
\usepackage{url}            % simple URL typesetting
\usepackage{booktabs}       % professional-quality tables
\usepackage{amsfonts}       % blackboard math symbols
\usepackage{nicefrac}       % compact symbols for 1/2, etc.
\usepackage{microtype}      % microtypography
\usepackage{lipsum}
\usepackage{fancyhdr}       % header
\usepackage{graphicx}       % graphics
\graphicspath{{media/}}     % organize your images and other figures under media/ folder

\begin{document}
\include{definitions_of_references}

\title{%
  The EL-X8 computer and the BOL detector\\
  \normalsize Networking, programming, time-sharing and data-handling\\
  in the Amsterdam nuclear research project `BOL'\\ 
  A personal historical review}

\author{Ren\'{e} van Dantzig \\
 National Institute for Subatomic Physics (Nikhef) \\
  rvd@nikhef.nl}

\maketitle

%%%% Cite as
%%%% Update your official citation here when published 
%\thanks{\textit{\underline{Citation}}: 
%\textbf{Authors. Title. Pages.... DOI:000000/11111.}}       

%\vspace{2 cm}
\begin{center}To appear in G. Alberts and J.F. Groote (Eds), \emph{Tales of Electrologica}.\\ 
London: Springer Nature, 2022, pp 123-152
\end{center}
\vspace{1,5 cm}

\begin{abstract}
\noindent
From 1967 to 1974, an Electrologica X8 computer was installed at
the Institute for Nuclear Research (IKO) in Amsterdam, primarily
for online and offline evaluation of experimental data, an application
quite different from its `brother's', X8's. During that time, the
nuclear detection system `BOL'\footnote{`BOL' is Dutch for `SPHERE'.} 
was in operation to study nuclear
reactions. The BOL detector embodied a new and bold concept. It
consisted of a large number of state-of-the-art detection units,
mounted in a spherical arrangement around a target in a beam of
nuclear particles. Two minicomputers performed data acquisition and
control of the experiment and supported online visual display of
acquired data. The X8 computer, networked with the minicomputers,
allowed fast high-level data processing and analysis. Pioneering
work in both experimental nuclear physics as well as in programming,
turned out to be a surprisingly good combination. For the network
with the X8 and the minicomputers, advanced software layers were
developed to efficiently and flexibly program extensive data handling.

\end{abstract}

% keywords can be removed

\section{Introduction and background}
\label{sec:dantzig1}
%Contents 
%1.	Introduction and background
%2.	The BOL project
%3.	Why and how the X8 came to IKO
%4.	Setting up the computing and network infrastructure
%5.	Time-sharing 
%6.	Developing data processing and analysis software
%7.	Software for minicomputer network nodes 
%8.	Final considerations and conclusions
%9.	Acknowledgements

The establishment of the Mathematical Center (MC)\footnote{The Mathematical Center
(MC) was founded in 1946 and housed in a school building in East Amsterdam. The first
director was Johannes van der Corput. The MC installed a Computation Department in
1947, led by Adriaan van Wijngaarden. In 1983 the name was changed to Center for
Mathematics and Informatics (CWI).} and of the Institute for Nuclear Research
(IKO)\footnote{IKO was established in 1946 in buildings of a former gas factory in
East Amsterdam. The founding partners were Philips, the Foundation for Fundamental
research on Matter (FOM) and the City of Amsterdam. The first director was Cornelis
J.~Bakker, who later became director of CERN, the institute currently known as
European Center of Particle Physics (Cern) in Geneva. In 1975, IKO became part of
NIKHEF, the National Institute for Nuclear- and High-Energy Physics -- later renamed
the National Institute for Subatomic Physics (Nikhef).} was firmly rooted in the
post-World War II revival of science in Europe. One of the many important
ramifications was the groundbreaking development of computers at the MC, in
conjunction with a large-scale nuclear detection system at IKO. As described in this
article, direct and indirect links between these two successful institutes proved to
be extremely fruitful, in particular for IKO.

It was in the summer of 1953, when I was 16 years old, that I had the opportunity to
spend six weeks in the computing department of the MC, where I was able to `assist'
in assembling and testing circuits for the second version of the ARRA
computer\footnote{This `job' was arranged for me by prof.~David van Dantzig,
mathematician, statistician, MC department head, one of the founders of the MC, and
my father, who paid me the equivalent salary of a simple holiday job at an enterprise
(my original intention), because he thought that staying at the MC would be better
for my future. He was right.}. There, I met prof. Adriaan van Wijngaarden, Bram
Loopstra, Carel Scholten, Edsger Dijkstra, and Jaap Zonneveld, the team responsible
for a decade of inventive computer developments\footnote{At the MC the first Dutch
computers were developed: i.e.\ vacuum tube machines ARRA-I (1952), ARRA-II (1954)
and FERTA (1955); later the partly transistorized ARMAC (1956) and the fully
transistorized X1.} at the MC, the early pioneers of Dutch information and computer
technology in hardware as well as in software. Their work was continued at the first
Dutch computer manufacturer, Electrologica\footnote{Electrologica was founded in 1956
by A.~van Wijngaarden, B.J.~Loopstra and C.S.~Scholten from the MC, and
J.~Engelfriet, from the assurance company Nillmij. Electrologica developed and
produced the X1 (1959) and finally the `eightfold' faster X8 (1965). In 1968
Electrologica was taken over by Philips and renamed `Philips-Electrologica'.}.

The initial series of Electrologica computers was the `EL-X1', the first of which was
installed at the MC in 1960. The MC offered courses on coding the X1 machine. These
courses were open to people from outside the MC, and I, at the time a physics student
at IKO, attended these, receiving a coding certificate in 1961. On top of that, Van
Wijngaarden's inspiring classes in numerical mathematics at the university of
Amsterdam made me and others at IKO familiar with programming techniques in the
Algol-60 language. This turned out to be quite relevant for IKO, since in the early
1960s the need for data reduction, analysis, and theoretical modeling increased
significantly.  For me, ist meant that my use of the very noisy mechanical Monroe and
Friden motor driven calculators was steadily replaced by silent programming. The MC
had created an open-shop service for the EL-X1 computer, which allowed us physicists
and technicians to write our programs during daytime and run these ourselves at the
MC during evenings and nights. We spent nights running programs on the X1 with a
portable radio on the memory cabinet: the beeping and crackling of the radio
interference kept telling us how far the data processing had progressed.

Our institute, IKO, had been set up to house a synchro-cyclotron\footnote{In
principle, a circular accelerator can reach much higher energies for charged
particles than a linear accelerator. In a synchro-cyclotron, the high frequency/high
voltage acceleration mechanism is adapted to the relativistic mass increase of
particles with increasing energy; \usereference{Heyn1952};
\usereference{Luijckx1980}.}, the first circular accelerator of nuclear particles in
Europe and at that time the second largest in the world. The cyclotron, based on
principles developed in the US, had been constructed by the Philips company. It
became operational in 1949, enabling irradiations inside the cyclotron. Accelerated
particles of various types\footnote{Originally only deuterons and alpha-particles
($^4$He-nuclei) could be accelerated, later also protons and $^3$He-nuclei.} with
various energies could hit a target inserted at the edge of the cyclotron. Nuclear
collisions produced radioactive nuclei inside the target, which was removed after
irradiation\footnote{In 1966, the 15,000$^\textrm{th}$ internal irradiation took
place.} for radiochemical and nuclear spectroscopy studies, as well as for medical
purposes.

Not long before I started as a student at IKO in 1959 the cyclotron had been equipped
with an external beam system to guide nuclear particles to a target far outside the
cyclotron, where they collided with the target nuclei. Generally, nuclear particles
can fly away from such collisions in `any' direction. When caught in special
detectors near the target in a vacuum scattering chamber, detected particles can
reveal aspects of the collision process, of the nuclear reactions that occur and of
the structure of the nuclei involved.

The IKO `nuclear reaction group', also called `deflection group', in which I
participated for the experimental part of my study, had been responsible for the
extraction and deflection of the beam from the cyclotron and its use for nuclear
reaction experiments. In addition to the internal irradiations, the external beam
with various instrumental setups made it possible to perform a range of nuclear
experiments, including the one for my own master's thesis in 1962.

In the first few years of the 1960s, a second accelerator was built at IKO, the
linear electron accelerator `EVA'. All the activities relating to both accelerators
were part of a broad and fruitful research program that was a rich source for
publications in international journals over the years. In this article, we will
concentrate on parts of the work with the cyclotron.

\section{The BOL project} 
\label{sec:dantzig2}
In 1964, I returned to the nuclear reactions group after a one-year stay
abroad\footnote{At the Weizmann Institute of Science, Rehovot, Israel, I studied
nuclear theory and made an analysis of my master's thesis measurements. This was made
possible by a stipend from the Dutch organization for scientific research (ZWO).}. By
that time, measurements at IKO as well as in laboratories elsewhere had shown the
importance of detecting two or more particles emerging simultaneously from the same
nuclear collision, called `coincidences'. These provided much more detailed
information on the ongoing nuclear reactions and on the structure of the colliding
nuclei compared to the more common `singles' measurements. This was particularly true
for `few body reactions', which could be well described theoretically and give
information about nuclear dynamics and forces.

After various feasibility studies, initiated by our group leader, Leo A.Ch.~Koerts, a
broad instrumental development was set up to construct a brand new type of detection
system to measure coincidences.\footnote{\usereference{Koerts1971}.} The system would
include a then revolutionarily shaped, spherical scattering chamber around the
target, containing many detection units, the whole resembling an inverted insect's
eye. This became the BOL project.\footnote{\usereference{Dantzig2017}.}

One underlying idea was that the number of pairs of detection units rises almost
quadratically with the number of detection units.  Therefore, also the probability to
catch coincident particles would rise quadratically with that number. Another
principal idea was that the detection units, when placed in a spherical arrangement,
could sample nearly the entire measurement space at the same time.

An ambitious team of enthusiastic physicists and technicians, the BOL team, was
formed around Koerts, the founding father and overall coordinator of the BOL project.
Three very motivated physicists in their twenties, Karel Mulder, Jona E.J.~Oberski,
and myself, coordinated the developments of, respectively, mechanical and detection
hardware, electronics and computing hardware, and software. However, all three of us
were more or less involved in all BOL activities. This meant that we had a broad
spectrum of unofficial duties and responsibilities, although none of us had a PhD
yet.  These were different times from today! We had a semi-permanent employment
contract at IKO and were not supposed to write our PhD theses until we had obtained
our own significant physics results, whatever time it would take.  Building the
equipment to obtain these required many years' work, including substantial delays!
Gradually, the team was expanded with younger PhD students, master's students and
technicians. Most of them made vital contributions to our project. From the
beginning, we operated in a remarkably free environment, based on mutual trust and
equality, without any authoritarian hierarchy or bureaucracy.

\begin{figure}[ht]
\begin{center}
\includegraphics[scale=.92]{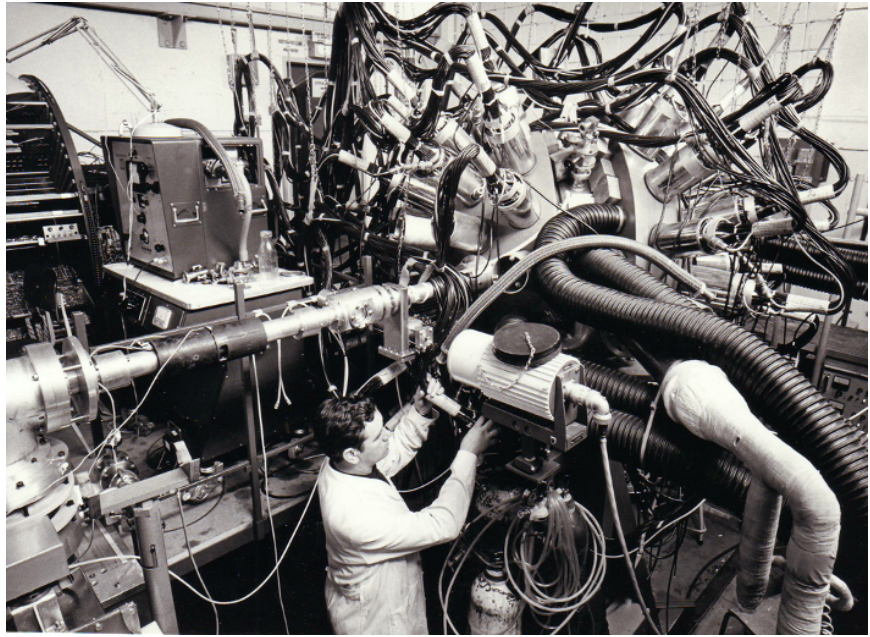}
\end{center}
\caption{The BOL detector, with technician Wil Verlegh at work. 
The particle beam can pass through the pipe coming from the left side}
\label{fig:dantzig1}
\end{figure}

The BOL detector was mainly designed and constructed during the years 1964-1966. 
The apparatus came into operation with a steadily growing number of rocket-shaped
detection units, ultimately 64 in all. When mounted, each had its silicon
semiconductor `detector telescope' in a vacuum scattering chamber, cooled at minus
$20^\circ$C and pointing at the target; the external part contained air-cooled
front-end electronics (Fig.~\ref{fig:dantzig1}).

BOL was the first `4$\pi$-type'\footnote{`4$\pi$-type' means sampling all 4$\pi$
steradians, the total maximum solid angle (the surface of a unit sphere) by an
internally directed omnidirectional looking fine-meshed `fly's eye'. The coverage for
detection was typically 10\% for single particles and 1\% for coincidences, values
experimentally unprecedented at that time.} nuclear physics multi-detector system in
the world that simultaneously covered largely all scattering angles and could
determine the `type' of the detected particles.  It was equipped with
state-of-the-art detectors developed in close collaboration with the BOL team by the
Philips NatLab Nuclear Detection group.\footnote{\usereference{Dantzig1980}. The
Philips NatLab Nuclear Detection group led by dr.~Wim Hofker included Piet Bakker,
Jarich Politiek, Dick Oosthoek, and others. It was housed at the IKO campus.} An
essential innovation was the Checkerboard detector, which made it possible to
determine a particle impact position with millimeter precision, corresponding to
typically 1 degree in scattering angle. This detector turned out to be a first step
in the development of detectors currently essential in high energy particle physics
experimentation.\footnote{\usereference{Heijne2003}.}

\begin{figure}[ht]
\begin{center}
\includegraphics[scale=.8]{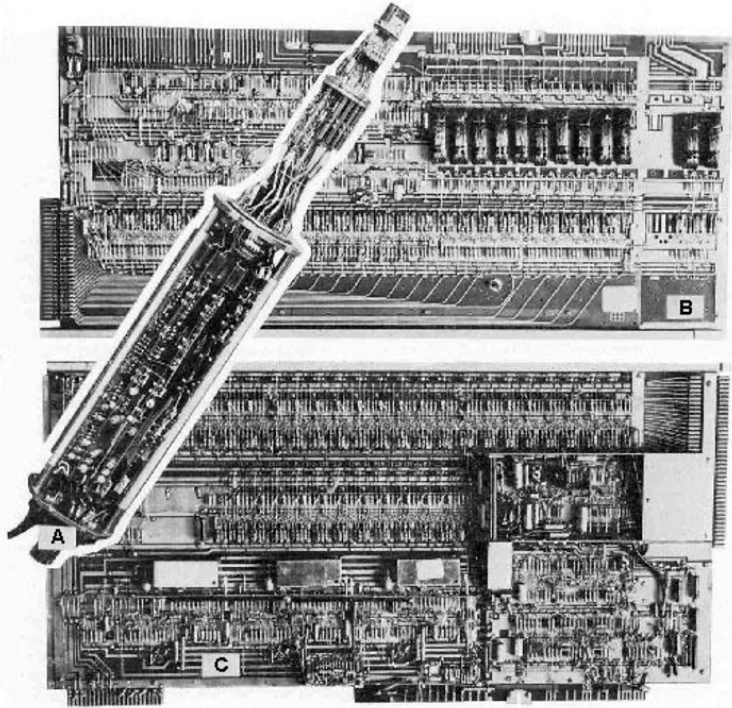}
\end{center}
\caption{BOL-electronics for one of the 64 detection channels: 
   [A] detection unit, from top to bottom: detector telescope, pre-amplifiers, 
       vacuum-air separating flange, processing electronics for the analog signals; 
   [B] printed circuit board that contains the logic and control electronics; 
   [C] printed circuit board that digitizes the detection signals.}
\label{fig:dantzig2}
\end{figure} 

In line with the complexity and the new detection methods, the technical realization
of BOL implied new approaches and specialized expertise\footnote{The electronics were
designed and developed with indispensable ingenuity and perseverance by Rein
F.~Rumphorst, head of the electronics group, assisted by Johan Dieperink, Erwin Kok
and others.} on the design, testing, and production of a massive and at the same time
delicate electronic system.\footnote{\usereference{Oberski1971a}.} Each of the 64
detection channels consisted of electronics for a) preprocessing the signals inside
the detection units, as well as two large printed circuit boards with b) system logic
and c) digitization using several analog-to-digital converters (ADCs). One of these
-- as a masterpiece -- was a 4096-channel high-precision highly linear ADC operating
at 100 MHz, a marvel far ahead of its time. The $2 \times 64$ boards
(Fig.~\ref{fig:dantzig2}) were radially mounted in adjacent cylindrical frames, the
`double drum'.  To illustrate that all the hard work was not without humor, each
ADC-board carried a motto, which freely translated to: `Computer at any speed, this
ADC will floor it indeed'.

\begin{figure}[ht]
\begin{center}
\includegraphics[scale=.87]{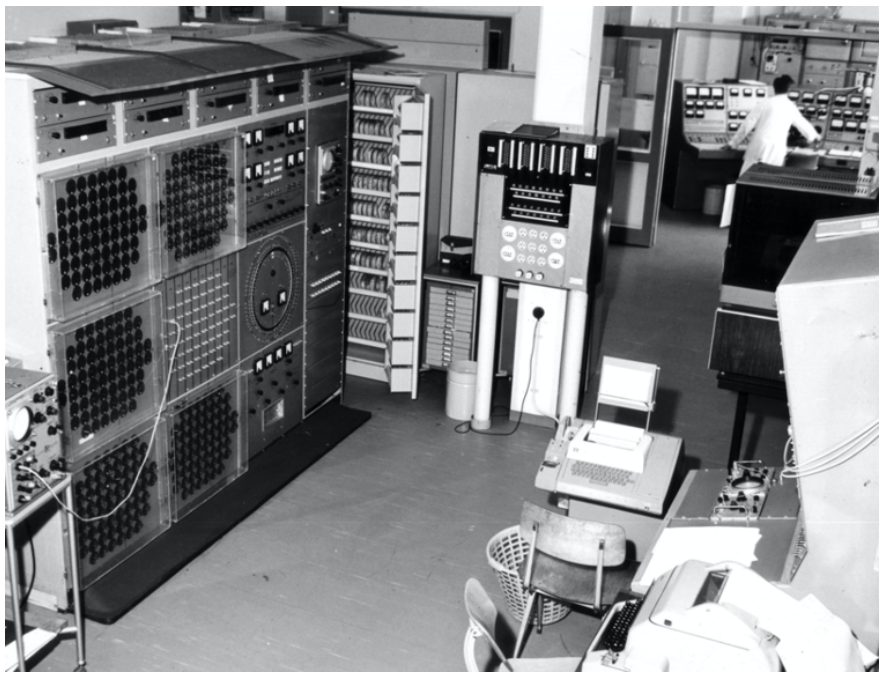}
\end{center}
\caption{BOL control panels at left, display station and PDP8-B with teletype at     
  right, and the cyclotron control console with the operator at the back}
\label{fig:dantzig3}
\end{figure}

Compared to what was available commercially those days, the resulting BOL electronics
were unrivaled in terms of precision, measurement speed, and size.\footnote{In total,
the BOL electronics contained about 40,000 transistors, 100,000 passive components,
35 m$^2$ circuit board, and 13 km of cabling, all of which were huge quantities at
that time for such equipment.} Each of the 64 detector units had its own electronic
channel with manual control of several voltage parameters (Fig.~\ref{fig:dantzig3}).
All channels had at the-back end a 72-bit register. This held all digital data of one
detected particle, a `single particle event'.  The 72 bits were made-up of `bites',
each containing a measured value -- for instance, an energy signal from the detection
telescope (12 bits), a similar signal from the checkerboard detector (7 bits), a mark
to indicate which single particle events belonged together to form a coincidence
event, and others. The `event data' were read out from the detection channels through
a sequencer by a minicomputer (see Section~\ref{sec:dantzig7}), part of a dedicated
computer network.

From 1965 onwards, the BOL detector and its infrastructure were gradually produced,
assembled and commissioned. Creating BOL gave rise to a variety of technological
challenges for IKO, and the project involved about half of the staff of the
institute. Their joint effort: assembling a complex mechanical construction with
subtle cooling and vacuum hardware, fragile detectors, and the intricate electronics
system. For everyone involved, these years offered a wealth of experience. A lot of
details had to be thought through in advance, while it was not yet clear how matters
would turn out.

In 1968, we obtained the first results with successively 10, 20 and 30 detection
units. From that time on, the system was systematically expanded, debugged and
increasingly used for experiments with day and night data recording periods and
subsequent analysis. This continued until the summer of 1973, when BOL was put on
stand-by.

Innovative as they were, BOL and the cyclotron had their limitations.  In order to
avoid overload by accidental coincidences, the BOL experiment could handle only a
low-intensity beam, matching the limited properties of the nearly two decades-old
cyclotron, which was by itself no longer competitive with the modern cyclotrons that
appeared elsewhere in Europe. For a decade BOL contributed significantly to the
useful life of the `aged' IKO-cyclotron, which was switched off in 1975.

\section{Why and how the X8 came to IKO}
\label{sec:dantzig3}

At the beginning of my work at IKO in 1960, before the BOL-project, nuclear
measurements were visualized online as 1-dimensional (1D) or 2-dimensional (2D)
distributions of the number of detection signals (pulses) as a function of their
height, `pulse height spectra'\footnote{The pulses were sorted according to their
height in bins and counted per bin.}.  These distributions were obtained with a
`kicksorter' or `pulse-height analyzer', which contained one or two ADCs coupled to
core-memory. In tandem with the recording process, a cycling process went along all
memory addresses and displayed the spectrum dynamically on a CRT (cathode ray tube)
display. Thus, we could visually follow during the experiment various spectra,
corresponding to detection signals. Whenever a typical spectrum reached sufficient
significance, we stopped the accumulation, studied the spectrum in detail and decided
if it was what we wanted. If not, the experiment was readjusted and the procedure was
repeated. The spectra could be printed, punched or saved on magnetic tape.

In general, a physics measurement with many adjustable elements in the system needs
continuous monitoring and supervision of the system behavior. In addition, at the
start-up of the measurement and at any time when changes are to be made, correct
parameter adjustments or re-adjustments depend on adequate feedback concerning the
system functions.

When we learned of small online computers in the US, it was immediately clear to us
that replacing a kicksorter with a small general-purpose computer would be of
tremendous advantage. The latter could do all of the data handling of the kicksorter
much better, with much more flexibility and capacity!

When Digital Equipment Corporation (DEC)'s PDP-8 minicomputer\footnote{This `Classic'
`Programmable Data Processor', PDP-8, was the first real minicomputer (1965) having
4K 12-bit memory (32 pages of 128 words) and direct addressing within a page. The
machine came with a Teletype ASR-33 serving as a console as well as paper tape punch
and reader. Our machine also had a DF32 fixed head disk system with a storage
capacity of 32k 12-bit words.  In this article, we will refer to PDP-8 as `PDP8'.}
came on the market, we immediately purchased one for the data
acquisition\footnote{\usereference{Dantzig1971a}.} in the BOL project.  We remember
the wooden crate that came by plane from the US, from which a gleaming
half-transparent PDP-8 computer emerged. All of us immediately fell in love with it!
A second one, also for BOL, followed soon; several more were bought for other IKO
research groups. From the beginning, perhaps because of our X1 coding experience,
some of us were fascinated by programming these machines. The PDP8 4K Disk Monitoring
System that came with them was primitive and not well-suited to our data acquisition
needs. I initiated several developments, some of which are discussed in
Section~\ref{sec:dantzig1}. Although both PDP8 machines offered us remarkable
possibilities and were certainly essential for data acquisition and for experiment
control, their capacity was far from sufficient for the needs we anticipated for
online data handling for BOL.

From 1962, at all of IKO, the steadily growing computing requirements were annually
evaluated and reported. A committee was appointed to prepare a plan on how to meet
the future calculation needs. It was anticipated that the necessities for experiments
at both IKO accelerators would be enough reason to seek funding for an in-house
general-purpose computer. Up to that time, the offline calculations had practically
all been done on the X1 at the MC. In 1964, it was foreseen that an online computer
would ultimately be needed in 1966 to cover all IKO computing work.

In December 1962, Lout Jonkers, already acquainted with many aspects of computing,
started his master's study at IKO on our in-house computer issues. In 1964, after his
first year of orientation and study, he wrote an internal
report\footnote{\usereference{Jonkers1964a}.} concerning computer requirements for
the various experimental setups in operation, in which he called for minicomputers as
a solution to the experimental problems. At that moment, it was not yet clear at the
laboratory level how fast new instrumentation (BOL) could change the requirements. In
a slightly later report,\footnote{\usereference{Jonkers1964c}.} he argued that: ``for
the sake of following the experiment and experimental control, the immediate
interpretation of preliminary results requires complicated computations and data
processing. Large quantities of information will have to be stored with high speed
and in such a manner that later offline data reduction by a computer can be done
economically.'' This report also addressed the need for ``multiprogramming with
hardware program relocation and multi-level indirect addressing'' and stated that
``it is highly desirable that there will be close cooperation between the
manufacturer and our laboratory.'' We fully agreed with all of these points, and his
discussions with us may have fleshed out the particulars.

The X8 in development at Electrologica had already been in the picture as a possible
future IKO-machine since 1963. An ad hoc
comparison\footnote{\usereference{Wapstra1964}; \usereference{Jonkers1964b}.} between
the X8 and the CDC 3200 from Control Data Corporation, a US company particularly
active in Europe, showed advantages\footnote{Although the X8 cycle time was longer,
the instruction set was considered considerably more powerful.} for the X8 -- partly
because of its appealing broader functionality and partly due to the possible support
of Electrologica on hardware and software development. For comparison, offers were
requested from several US computer manufactories, including CDC. It turned out that
only the least expensive machine, the CDC 3200, was at all affordable. In 1964, the
X8 -- almost ready for rent or lease -- became the anticipated (dreamed) machine for
BOL as well as for other IKO applications, and consultations took place.

On the basis of the above evaluations, in 1964, IKO requested funding for the fiscal
year 1965 from the overarching research agencies FOM (Foundation for Fundamental
Research on Matter) and ZWO (the Dutch organization for pure scientific research) for
the procurement of an electronic computer.\footnote{\usereference{Wapstra1965}.}
Almost at the same time, the high-energy physics (HEF) community at the Zeeman
Laboratory in Amsterdam requested funding along the same route for a computer, online
and offline, for equipment measuring high-energy particle tracks on pictures taken at
Cern. Unfortunately for us, the IKO request for 1965 was not approved; that year,
high-energy physics was prioritized\footnote{The competing demands in 1966 for
expensive computing facilities from the side of the high energy physics (HEF)
community as well as from IKO, the nuclear physics institute, prompted FOM to propose
to ZWO to combine both in one national institute of nuclear and high-energy physics.
However, it was not until almost a decade later, in 1975, that this institute, NIKHEF
was created, where at that moment my employment contract with FOM was continued.}
with a rented CDC 3200.

Still, at the same time, the online coupling of the computer to BOL became the
spearhead of an updated funding request specifically for the X8, with convincing
arguments being made for the X8, in comparison with the foreign (US) alternatives.
This time, the request was honored! Delivery would be in 1966; it actually turned out
to be 1967, two years later than we had originally planned. On the other hand, such
delays also occurred in other parts of our project, so we had gotten used to this.

\section{Setting up the computing and network infrastructure}
\label{sec:dantzig4}

As mentioned, before online computers were available, we checked and adjusted the
performance of our experimental setups using `measured data' from an online
kicksorter on an oscilloscope display.  We realized that when going to work with 
BOL, we were bound to change our habits and base our adjustment decisions on more
significant `preliminarily analyzed data', to be delivered by a computer. In this
spirit, hardware and software for the approved EL X8 were prepared.  In 1965, the
contract was signed between IKO/FOM and Electrologica, in which it was agreed that 
in addition to the standard computer with various peripherals, a hardware buffer
interface, `IKOB', with accompanying software and several software packages would 
be supplied by Electrologica according to mutually agreed specifications.

In conjunction with Electrologica staff and together with my collaborators, I had
interrupted my involvement on the PDP8s (see Section~\ref{sec:dantzig7}) in order 
to specify the X8 packages. Similarly, my close collaborator, Jona Oberski, in
consultation\footnote{In particular with Philip Seligmann from Electrologica.} with
Electrologica, came up with the specifications for the network communication between
the X8 and the PDP8s. None of us had experience with large software packages or with
networking for an X8, but that was also true for most of the BOL-developments: the
learning curves were steep. We were told by Electrologica, that the communication
principles and design specifications for IKOB had to be formulated by us within three
months; otherwise, the X8 delivery would be substantially delayed.  Despite
Electrologica's expectation that this would be impossible for us, the deadline was
met! Perhaps that is why Electrologica could not meet the scheduled delivery.

In planning for the data processing and data analysis, we realized that BOL would
produce large amounts of individual, non-reproducible events, for each of which a set
of quantities would be measured simultaneously. On these raw event measurements, a
range of criteria and conversions would have to be applied. Ultimately, not these
event data themselves would be the final goal, but rather a `bundled' representation
of them. Additionally, such bundled data, `spectra' in a broad sense, needed to be
processed further, in order to obtain the final results of the experiment that
allowed physics interpretations.

Concerning the network communication, our point of departure was that the burden of
the `clerical control' should primarily be carried by the X8, since that machine was
equipped with sophisticated input/output (I/O) facilities. On the other hand, a PDP8
machine incorporated in the network would have to give final permission before any
data transfer, even if it had taken the initiative. In the box in
Fig.~\ref{fig:dantzigBox1}, we give a synopsis of the network communication
guidelines.\footnote{\usereference{Toenbreker1965}; \usereference{Oberski1965}.}

To help carry over the data, the X8 contained a dedicated micro-programmable
I/O-processor, called Charon. Although the Greek mythological ferryman Charon would
row his passengers across the Styx to the other world himself, the X8-Charon could
delegate the actual `rowing' between computers to IKOB together with the fast
direct-memory-access (DMA) mechanism of the X8 (fast-channel-selector, SKK) and of
the PDP8s (data-break).

\begin{figure}[ht]
\begin{center}
\includegraphics[scale=1.0]{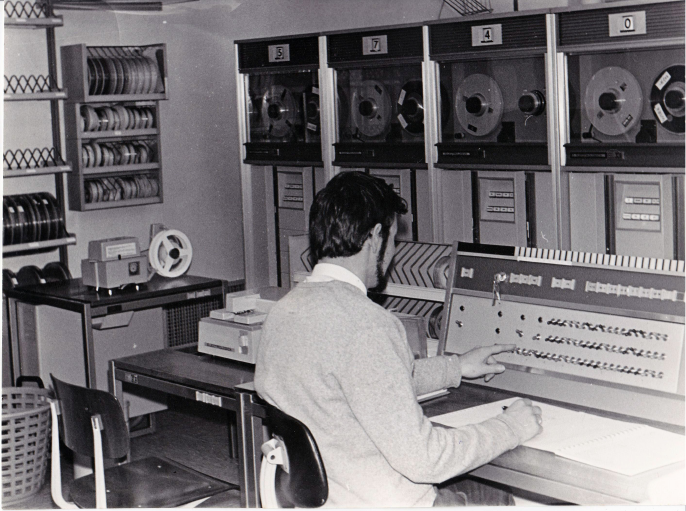}
\end{center}
\caption{The X8-room with programmer Anton Mars. From right to left: 
X8-console, magnetic tape drives, punched paper tape boxes, paper tape reader and 
  paper tape punch}

\label{fig:dantzig4}
\end{figure}

\begin{figure}[ht]
\begin{center}
\includegraphics[scale=.8]{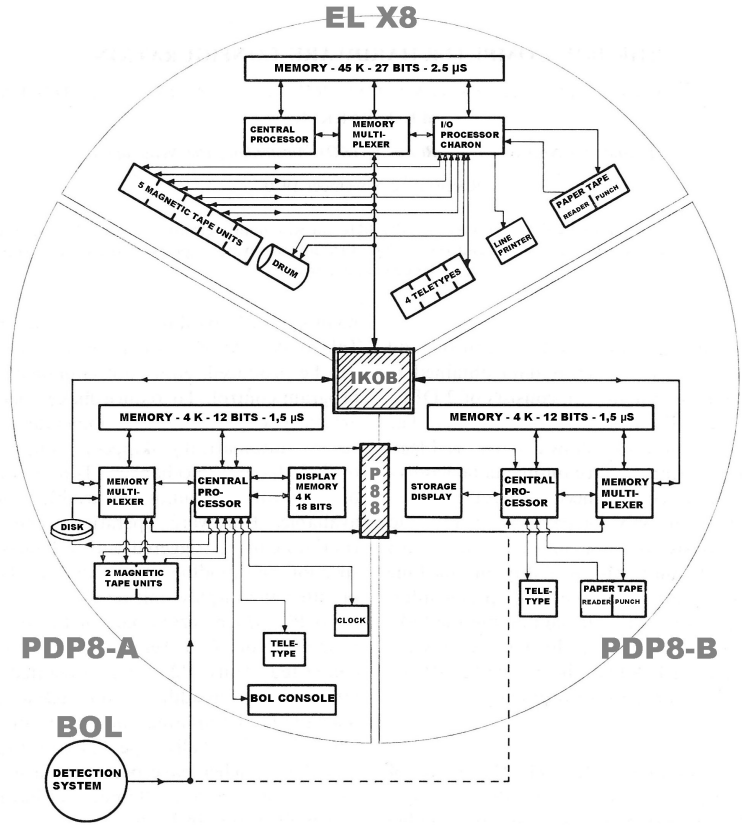}
\end{center}
\caption{Diagram of the triangular computer network realized for the BOL-system}
\label{fig:dantzig5}
\end{figure}

The installation of our X8 configuration\footnote{The X8-configuration consisted of
the basic unit with central processor (CRO), operator console, magnetic core memory
32k (later upgraded to 48k) 27-bit words (+parity bit) with cycle time 2.5$\mu$s, a
memory access multiplexer, Charon -- the I/O processor, fast channel selector (SKK),
Bryant magnetic drum with 512k memory of 27-bit words and a cycle time of 20 ms, real
time clock/alarm clock, console for manual control, Siemens teletypes, line printer
1-5 lines/s and for paper tape: a 1000 ch/s EL1000 reader and a 150 ch/s Facit
punch.} started in late 1966 and was completed at the beginning of 1967
(Fig.~\ref{fig:dantzig4}). Thereafter, the machine was also integrated into the
triangular BOL network with IKOB and the two PDP8s connected by the homemade
interface P88 (Fig.~\ref{fig:dantzig5}).\footnote{\usereference{Oberski1971a};
\usereference{Oberski1967}.} Shortly after, measurements with BOL were sent to the
X8s and secondary data coming from the X8 appeared on one of the display units with
which the PDP8s were equipped. From that time onward, adequate insight into the
experimental situation and the quality of the measurements of the functioning
BOL-system was available to the physicist overseeing the experiment.  Our coupling of
a complex nuclear physics detection setup to a computer network for the on-line
control and data handling may well have been a European first.

The next three sections describe aspects of the innovations we designed or developed
at IKO, partly in collaboration with Electrologica, on the computer side of our
installation: the time-sharing operating system and the X8 and PDP8 software
essential for the data-handling and analysis.\footnote{\usereference{Dantzig1971b}.}

\begin{figure}[ht]
\framebox{
\begin{minipage}{0.98\linewidth}
The communication between the X8 and a PDP8 is to be set up as a series of block
transfers, denoted `conversations', that can be initiated by any of the three
machines. Two types of sentences can occur in the conversations: `introductory
sentences' and `main sentences'.  An introductory sentence is a short data block of
fixed length and with a fixed format, providing information about at most one
following main sentence. Main sentences have a variable length and structure.  Charon
always has at its disposal addresses in memory space, where current sections for
input or output are located in all three machines. These memory sections are called
`repositories'.

\hspace*{0.3cm}If a machine wants to communicate with another machine by sending an
introductory sentence, it offers a communication request to Charon, specifying the
addressed machine. Charon then checks whether IKOB is `not busy' and -- if so --
investigates whether the addressed machine gives `input permission', implying that
this machine has an input repository ready to receive a block transport. Then Charon
arranges a data transfer through IKOB from the `output repository' of the source
machine to the `input repository' of the target machine. The PDP8s have fixed
repositories for input and output of introductory sentences. A flag (flip-flop/bit)
in IKOB indicates whether the input repository is available. Transfers always involve
an even number of PDP8-words of 12 bits being packed in 24 bits, on the X8-side
fitted into a 27-bit word.

\hspace*{0.3cm}The repositories in the X8 are chains of relocatable memory sections,
connected by address pointers and provided with chain counters.  When a subsequent
request comes before the current repository is available, the next section in the
chain is used. Upon a communication request, Charon reads the address of the
repository and arranges the transfer. When the introductory sentence has been
transferred in DMA-mode, with a word transfer counter in IKOB, the latter informs
Charon with a dedicated flag.

\hspace*{0.3cm}Also, if the X8 takes the initiative, a main sentence is handled by
Charon only following a request from a PDP8. In that case, the PDP8 is `invited to
invite' Charon to send an introductory sentence with the required PDP8 details
included. When a main sentence DMA transfer has been completed, IKOB notes this in
this case, too, and informs Charon with a dedicated flag.

\hspace*{0.3cm}Because, on the PDP8 (experiment) side, the available memory is
restricted and speed is important, a single main sentence can be transferred in
`stutter-mode', implying that the transfer is split up in parts separated by a pause,
such that the PDP8 can use the same memory section a number of times, thus saving
space. The stutter mode does not allow the transfers to be checked rigorously and
thus cannot be used if even an exceptional error cannot be tolerated, such as with
transfers of programs. For measurement data from BOL, this mode is considered
acceptable.

\hspace*{0.3cm}The simplest and fastest mode is `unsolicited reading' of PDP8 memory
by the X8. In this case, as long as IKOB is `not busy', the X8 can read a specific
part of PDP8 memory (a section with measurement data) and is supposed to know how to
handle the obtained data.

\hspace*{0.3cm}Software drivers for IKOB are needed on both types of machines in such
a way that data transfers satisfy standards already existing for the data streams to
and from the magnetic tapes. This is important in order to facilitate data handling
procedures independent of the specific input and output channels.

\end{minipage}}
\caption{Summary of the PDP8 $\leftrightarrow$ X8 $\leftrightarrow$ PDP8 network 
  communication guidelines / requirements, in accordance with informal reports by 
  Toenbreker (1965) and Oberski (1965)}
\label{fig:dantzigBox1}
\end{figure}

\section{Time-sharing}
\label{sec:dantzig5}
Almost all computers in the 1960s relied on `batch processing' or `unitasking', where
one job is executed after the other. `Time-sharing' (TS)\footnote{Time-sharing
implies `multitasking' and `multiprogramming', but the converse is not necessarily
true. In the latter cases, jobs submitted in batch-mode can overlap in time and in
system resources, but there is not necessarily a possibility for multiple users
working parallel to each other. The first Dutch application here was the
`THE'-multiprogramming system developed by E.W. Dijkstra, where jobs are submitted in
batch mode, but run as much as possible interleaved, to gain overall runtime.}, where
a number of users can work in parallel, each of them having access to essentially all
system resources, had been developed in the US. Worldwide, the first TS system became
fully operational in 1961 at MIT in Cambridge. A little later, several TS systems in
the US were described in the literature, but -- as far as we know -- no such systems
were available in Europe at that time.

As installed, the X8 operating system, called `Monotor', provided only batch
processing. When, in the course of 1967, the system became intensively used, the
shortcomings of batch processing were felt more and more. Often, time was lost with a
change of user.  Immediate high priority access to the computer was rarely possible,
and conversational control of programs was hardly allowed, because it was considered
inefficient and it would preclude access by another user for quite some time.

Fortunately, within one year, Pieter van Engen and Rolf Meesters of the IKO Software
Group\footnote{The IKO software group led by Pieter van Engen included Anton Mars,
Rolf Meesters and Pim Biekman.} developed a time-sharing system, called
`Wammes',\footnote{\usereference{Engen1969}.} which included a few parts of Monotor,
such as magnetic tape drivers. On starting this project, the authors were inspired by
the work of Edsger W. Dijkstra\footnote{\usereference{Dijkstra1968}.} on `semaphores'
and `mutual exclusion'\footnote{Mutual exclusion for concurrent processes has been
studied by Th.J.~Dekker. to safely guide concurrent processes through`critical
sections' that had the risk of -- for example -- a `deadlock'.}

Wammes became fully operational in March 1969 and allowed users to load, assemble,
execute, trace, stop and inspect programs simultaneously at each of the four
terminals. The system and the programs it could handle were written in the X8
assembler macro-code ELAN.  The assembly of a program resulted in a binary
relocatable\footnote{Upon assembly, the resulting machine program was, using a simple
trick, equipped with `relocation information' to be used at start-up time.} file,
which could be appended to the system library, including the program text.  In the
absence of discs, the file system was a mixed system of the drum and magnetic tapes.
A user could create a file on the drum and dump it onto the central library, a
magnetic tape, from which it could be retrieved later.

The system was based on a carefully operating swapping method. If possible, a program
was loaded in an unoccupied core area or elsewhere on the drum. When a program in
core memory was blocked, only the part that was needed for another program was
swapped. Except for magnetic tape units, which a program had to reserve in advance
for its own use, all other programs could use every input/output device virtually
simultaneously. This was achieved by buffering the I/O data on the drum. A driver
program corresponded to any output-device, selecting the next `closed' file for
actual output. Every program could use dynamic storage allocation routines which
operated in the working space of the program.

One of the benefits was that the time necessary for programming was considerably
reduced due to the editing, debugging and library facilities, making more time
available to more users. The possibility to split and run a program in segments, as
we developed for the PDP8 (see Section \ref{sec:dantzig7}), and as often done in
other TS systems, was not possible for the X8. Every program could only be relocated
in its entirety.

Priorities and machine time allocation to logged-in users were essential. The
starting point was the democratic principle that all users get the highest priority,
and in turn, the same slice of processor time. However, there were a few other rules,
too. For example, the response time at terminals was -- as much as possible -- within
seconds. When a program was involved in semi-real-time data processing for the BOL
experiment, it had `urgency priority', which meant that it got all the processor time
it needed and could get and would not be swapped out, unless manually by the
operator.  If there were two urgent programs, the first one received the highest
priority.

An important software debugging facility, which came to full fruition in Wammes, was
`dynamic program tracing'. It was available on our X8 exclusively, and originated
from one of the early meetings with Electrologica, where we reported our positive
experience with PDP8 program debugging using our own development of `tracing' under
dynamic supervision. We had discovered that a very simple trick made this possible
and we asked Electrologica whether something similar could be done for the X8. The
method, which we called `delayed self-interrupt' (see Section~\ref{sec:dantzig7}),
imposed only extremely simple requirements on the hardware. The program, while
running, could be interlaced in time by a tracer-interrupt routine, operating in the
background of the program. It turned out that a similar trick was also possible for
the X8 and a program to be tested could run under the control of the programmable
tracer, while after the execution of essentially any machine
instruction,\footnote{Only instructions executed within a system program, where the
X8-interrupt was switched off, could not be traced, obviously.} the status of
registers and addresses could be checked on the basis of flexible criteria in the
tracer interruption routine. In a report\footnote{Note added to a brief Electrologica
report by Th.R.C.~Bonnema of a meeting end 1965, with present: Nossbaum, Seligmann,
Bonnema from Electrologica, and Koerts, Van Dantzig and Oberski from IKO, with the
title translated as: `Coordination facilities for a tracer'.} of a BOL-Electrologica
meeting, it was written that ``This tracing facility would in the first instance be
applied especially for IKO''.

Wammes was a `bomb-proof' time-sharing system: all user programs ran in a mode where
it was impossible to write outside their own memory space. This was realized thanks
to X8's top-of-the-line, hardware feature, write-protection to be set and cleared for
blocks of 512 words by system routines running in kernel mode.  Even if a user got
stuck in a loop, there was a clock interrupt at the end of the current time slice to
finish the user task.

The Wammes developers promised a bottle of good wine to the user who succeeded in
`hanging' Wammes. JEP de Bie, one of the physicists/programmers, accomplished this
the next day with a program having a single instruction: `Program Boom; Loop: do
Loop; end Boom'. The `do' command executed the instruction indicated in the address
field -- in this case, this instruction itself. Thus a single instruction was trapped
in an endless recursive loop, the system being immune for any clock interrupt. As a
result, the `blinking lights' on the console glowed continuously. The only recourse
was to hit the on/off switch and do a full reboot. Apparently, this possibility had
been overlooked in the hardware design by Electrologica. JEP de Bie thus got his
bottle of wine. I am not certain whether Electrologica corrected this.  Anyway,
deliberately hanging Wammes was rewarded only once; it became banned and no such
anomaly ever showed up in normal user programs.

When we consider Wammes from the user's point of view, in principle we could start,
pause and stop programs on the teletype as if the entire X8 were freely at our
disposal. Mostly, that worked out well, although it could happen that the system
refused a program because it would need too much drum memory or tape units, some of
these having already been reserved by other users. Basic characteristics of
time-sharing systems, such as the parallel use of peripherals by different users and
the simultaneous use of both the foreground and a background memory, were clearly
present. The goal, to build a system that could efficiently run essentially the same
programs as in batch mode, had been realized. The X8 with Wammes became a more
pleasant and better manageable machine.

\section{Developing data reduction and analysis software} 
\label{sec:dantzig6}
Soon, we became strongly committed to the design of an overall data reduction system
that would be as simple as possible, and could flexibly be programmed without
complications from I/O. To achieve our goals, we would `mold' our event data into a
universal structure according to a set of rules, the `BOL format'. The main idea was
that the data structure would have `levels' in substructure and that operations at
the various levels could be programmed separately, independent from where they came
and where they would go. Once in this BOL-format, the events could be subjected to
several user-programmed `passes' with checks, alterations, rejections, selections,
and calibrations, ending up on tapes as experimental sets of `validated data'.

All data were supposed to remain in accordance with the format during all the
processing. The BOL format had seven levels of nested data subsets, `quanta' of
information.\footnote{\usereference{Dantzig1967a}.} For primary BOL data at the
deepest level, the elementary quantum was a `record', specifically, an `event record'
that contained data from a measurement of one or more detected particles resulting
from one collision. At the next level, and only for practical
reasons,\footnote{Memory buffering and writing on external media like magnetic tape.}
a number of records were packed in a `block' (without a particular meaning). A series
of records (in blocks) would combine into a `train'; trains into a `net'; nets into a
`group'; groups into a `file'; and finally, files into a `repfile'. Trains, nets,
groups and files were separated by special blocks. Different quanta were meant to
have a specific meaning. For instance, a train (of blocks of records) might contain
homogeneous data, measured under exactly the same experimental conditions. A net (of
trains of blocks of records) might correspond to the same target with different beam
conditions. The largest quantum at the highest level was a `repfile', a series of
files, representing all data from one full experiment.

To obtain an overview and analysis of what was measured, we would have to `bundle'
validated event data. Bundling meant that certain combinations of values derived from
the event data were `sorted and counted' on a multi-dimensional (mD) grid in order to
produce mD `spectral arrays',\footnote{A train represented a 1D array, that was a row
of array records (0D-arrays), a net, a series of trains (1D arrays, rows, or a single
2D array), a group, a series of nets (2D arrays, or a single 3D array), and a file, a
series of groups (3D arrays, or a single 4D array).} `spectra' for short. A spectrum
was characterized by the numbers and the widths of the mD grid cells, together with
the number of counts collected in each cell. The coordinates corresponding to the
central values of the grid cells represented derived `physics values'.\footnote{For
experimental data, the `physics values' are derived from measured values.  For all
experiments it is common to produce event data by a Monte Carlo (MC) random sampling
method according to a theoretical model and all experimental effects taken into
account.  These MC-data are handled like experimental data and thus also can deliver
`theoretical values', to be compared directly with their experimental counterparts.}

The BOL format was defined in such a way that it could be used for spectral arrays as
well, to make these suitable for our standard data processing. An array element then
became an `array record', a row of the array became a train, a column of rows became
a net, and so on. A file thus could embody a series of 3D-arrays or a 4D-array.  For
event data, the sequence of quanta within their next higher quantum was irrelevant,
while for array data, this sequence was fixed by the positions (coordinates) in the
array.\footnote{Therefore, `array quanta' to be rejected could not be removed, but at
most administratively invalidated.}

In that same design phase, we realized that for our anticipated measurement accuracy
($0.1$-$1\%$), the size of mD arrays, in whatever representation, would be far too
large to be stored in their entirety in the X8 core memory. However, we could use the
drum memory, which was 16 times larger! It was helpful that the number of counts in
large spectral arrays would often be low in average and distributed in-homogeneously,
crowded in relatively small regions (peaks or mountain ranges) and almost or
completely empty in large regions, partly even by definition.\footnote{In many
spectral arrays, the kinematics of the reaction restricts the coordinate regions that
can be occupied by event counts at all.} High resolution would be only relevant in
regions with good statistics. With this in mind, together with Electrologica, we made
specifications for array-handling on drum store, allowing declaration and
manipulation of groups of flexibly sized 3D-arrays with array elements of 3, 6, 12 or
24-bit bites. In cases where small bites would be chosen, a certain percentage of the
array elements was allowed to `overflow'. This was taken care of by successively
assigning larger bites in a reserved overflow area. 

As opposed to the basic distributive\footnote{In general, the most compact build-up
and storage of spectra can be `distributive' or `associative'.  In the distributive
-- conventional -- case, the memory space for the spectrum is completely and
regularly structured beforehand such that all cells can be addressed directly given
the coordinates.} structure, the storage of these elements was
associative.\footnote{In associative storage, the structure in memory space is formed
during the build-up and depends on details of the spectrum. Then, the coordinates of
a cell are stored in an associated manner, i.e., together with -- or pointing at --
the cell content.} At read-out, these cells were to be picked up together with the
non-overflow elements in order to form the whole spectrum. A smart buffering and
synchronization mechanism\footnote{The method amounted to a) temporarily keeping the
coordinates of the addressed array elements in core memory buffers that were
associatively sorted according to drum tracks; b) asynchronous transport of drum
tracks to core memory and vice versa, giving highest priority to the tracks with most
modifications. Access rates of 104 per second were achieved.} was necessary to read
and modify the content of the drum addresses in an optimally efficient way.

Based on the above ideas, Electrologica started to build the corresponding software
layer, a package of subroutines that would take the format and structure rules as
given and provide the appropriate basic functionality for the data processing for all
future BOL experiments. This layer was called the `Window system' for the X8 (we here
call it `WSX'),\footnote{In Section~\ref{sec:dantzig7}, a simplified Window system
for the PDP8 (WSP) is introduced.} and the specifications were finalized with
Electrologica in 1966.\footnote{\usereference{Dantzig1967b}.}

The Window system with the drum array layer including documentation became
available\footnote{WSX was realized by Stef Toenbreker, Pieter G.~van Engen and Hans
Suys from Electrologica.} in 1967.\footnote{\usereference{Dantzig1968b}.} From that
moment on, it was central in our data processing. In the user programs (`passes'),
data could easily be split, switched, and mixed between different I/O channels, and
the programming was concentrated in `windows' at separate structure levels, dealing
with the various information `quanta' described. The data at a given level would pass
along the program window for that level. Program windows would only be opened for
levels that were relevant in the current processing. A program window could read,
alter or reject a quantum of data before sending it to one or more output channels.
ELAN was used to write both the system routines and the user's programs. From that
moment on, essentially all BOL data-handling was programmed and run in WSX.

Even with the above-mentioned arrays on the drum, we thought that we would be unable
to reach our overall required ultimate resolution, not even for only 2D array
spectra. Therefore, together with Electrologica we made specifications for an
additional package, called `counting in trees' or `associative counting'. In this
approach, 2D-trees, with a resolution consistent with a $4096 \times 4096$ grid,
would be built up beginning in core memory and continuing on drum storage. The tree
structure was defined so that each leaf of a tree contained an end node or maximally
4 pointers (2 for each dimension) to other leaves, each pointer corresponding to the
value of a certain bit in both coordinates.

Since each coordinate had a maximum of 12 bits, at most 12 pointer levels would be
needed to reach non-zero counts cells. In principle, an arbitrary number (maximally
32) of trees could be built simultaneously, restricted by available drum space. It
was, all in all, a quite complex framework. A buffer and synchronization technique,
algorithmically even smarter than for the WSX-arrays on drum storage, was made. The
package was delivered by Philips-Electrologica\footnote{In 1968, Electrologica was
taken over by the Philips company. } in 1969,\footnote{\usereference{Pouls1969}.} but
there were challenges to its use. The package was large, complicated to handle and
slow. Moreover, at that time we could already solve our problems with the available
software and the required resolution was less than anticipated. If my memory is
correct, despite the good work, this package may have been an instance where our
ambitions were too high.

Within WSX, spectra could be sent to an output medium, or immediately further
analyzed. The spectra could have peaks, valleys, flat parts, and empty parts, all of
these together constituting the `spectral structure' for the chosen parameters and
the chosen grid. Here, for the first time, we could judge whether the measurements
looked comprehensible, and hopefully correct. If not, we had to go back to the data
conversion passes and look for a possible explanation.  When the validated data
looked fine, the next crucial step could be taken, the identification of the detected
particles (see the box in Fig.~\ref{fig:dantzigBox2}).

\begin{figure}[t]
\framebox{
\begin{minipage}{0.98\linewidth}
Measurements from the BOL detection telescopes had been calibrated and converted to
the parameter values, $E$ (detected particle energy) and $\Delta E$ (energy loss in
the checkerboard detector). These parameters were characteristic for whether a
detected particle was of the type proton, deuteron, triton, $\tau$ (He-3) or $\alpha$
(He-4) particle. In an $E \times \Delta E$ versus $\Delta E$ spectrum approximately
parallel mountain ranges would appear, each range -- if present -- belonging to one
of the particle types. The particle identification required the contours of the
mountain ranges to be outlined and then made available to a new pass of event data.
There, using these contours, we could determine for each event the type(s) of the
detected particle(s), including the type: `unknown', in case the particle(s) fell
outside all contours. The unknown type could sometimes be resolved later using other
information. From this point onward, physics definitely came into the picture!
Suddenly, we could select a particular nuclear reaction process and study that in
detail, by now bundling the data to the most meaningful representations.  These were
real physics results, energy spectra, angular distributions or correlations, and the
like for a specific nuclear reaction. They could be interpreted, studied and compared
with theoretical model predictions. A selection of these results would appear in our
publications.
\end{minipage}}
\caption{Particle identification, an essential step in the data analysis}
\label{fig:dantzigBox2}
\end{figure}

In our team, a conversational program language, `Simplex',\footnote{Simplex was
developed by Bob J.~Wielinga from the BOL team.} influenced by Algol-60 had been
developed for calculations within the WSX framework. This was particularly important
in the final stage of the BOL data reduction and data presentation, where we wanted
to easily and flexibly apply our instrumental and physics insights to prepare
pictures for further study or for publication. Spectra could be presented on a
graphics storage display with hard-copy read-out connected to a PDP8 (see
Section~\ref{sec:dantzig7}). A useful extension of Simplex was the Lisp-based program
Lisi,\footnote{Lisi was developed by Theo F.~de Ridder from the BOL team, who had
installed and extended the LISP program of Van der Poel and Van der Mey.} which
provided facilities for the graphics display with formula manipulation and symbolic
processing of complex display structures.\footnote{\usereference{Ridder1972}.}

\begin{figure}[ht]
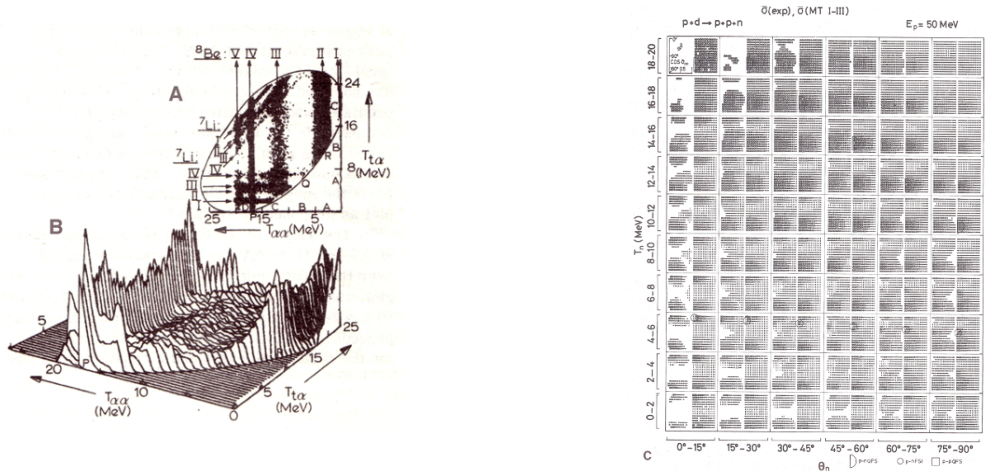

\begin{center}
\begin{minipage}{0.47\linewidth}
  {\includegraphics[scale=.7]{pictures/DantzigFigure7.pdf}}
\end{minipage}
\hspace*{0.5cm}
\begin{minipage}{0.47\linewidth}
  {\includegraphics[scale=.62]{pictures/DantzigFigure7a.pdf}}
\end{minipage}
\caption{Examples of published measurements in 2 and 4 dimensions. 
  [A]: 2D-topview of the measured reaction probability above a certain threshold for 
    the reaction $^9\mathrm{Be}+d\rightarrow t+\alpha+\alpha$, where the energies and 
    directions of all three particles in the final state were deduced from the data; 
  [B]: the same 2D data in a 3D isomeric display; 
  [C]: comparison between a measurement and the corresponding theoretical simulation   
     in 4D for the reaction $p+d\rightarrow p+p+n$. The bin-averaged data are shown 
     as a ($10 \times 12$) 2D array of a double ($15 \times 15$) 2D array (data, theory).
     The reaction probability is represented by special symbols for values in a 10-bin 
     range. Only the lower half range in $\theta_n$, that means half of the data, is 
     here shown}
\label{fig:dantzig7}
\end{center}
\end{figure}

For their physics interpretation, BOL results needed to be compared with theoretical
model predictions. Mostly this meant that, assuming the validity of the model, we had
to perform a full Monte Carlo computer simulation of the experiment, taking into
account the detailed measurement situation, including all the mounted and correctly
working detection units. Because of rotational symmetry around the beam axis, there
was a lot of measurement redundancy, largely compensating for missing detection units
(see Section~\ref{sec:dantzig8}).  When considered individually, our data were often
less accurate, both statistically and systematically, compared to measurements
performed elsewhere with fewer detectors. However, they covered almost all of the
kinematical final states of the nuclear reaction under study in a continuous way. And
that was unique! As foreseen, this was particularly important for the study of `few
body reactions'. A discrepancy with simulated data would indicate `that' -- and often
`where' -- the experiment showed deviations from the theory. A notable example was a
comparison in full 4D between an experiment and a theory simulation
(Fig.~\ref{fig:dantzig7}).\footnote{\usereference{Blommestijn1981}.}

The results obtained might clarify a known problem, but often new questions arose,
leading to a follow-up experiment to shed light on the new questions. For the
planning of new experiments, a crucial program\footnote{This program was developed by
Ton Ypenberg from the BOL team.} in Algol-60 was developed, applying relativistic
kinematics for nuclear reactions.

\section{Software for minicomputer network nodes}
\label{sec:dantzig7}
The hardware and software functionality of the two network PDP8s (A and B) was very
relevant in the X8 context. Our program developments for the PDP8s in the assembly
language PAL/8 were of importance for choices in the X8 software and vice versa.
Since interrupt handling was primitive for the PDP8 and we were starting on a basic
programming level, in the interaction with peripherals, we worked with `waiting' for
a ready `flag'\footnote{A flag was a `flip-flop' (bit) that could be set (1) and
reset (0) externally and/or programmatically.  It could be sensed by a
PDP8-instruction: if flag is set, skip next instruction. When properly programmed, it
could also give a program interrupt.} of each equipment, with one exception: tracing.
An operating system, called Monikor, was built by us, physicists, around a linked
list, containing all active waiting points and the corresponding values in the
accumulator register, which also had to be restored as soon as the program was
continued. Thus a compact and primitive multitasking system was created, without the
use of interrupts, which -- with later I/O extensions -- transferred all measurement
data from the BOL system for many years to magnetic tape and/or to our X8.

The Monikor system enabled dynamic storage allocation, binary program allocation,
multi-job processing with time sharing and memory sharing, asynchronous data streams
with semaphore mechanisms, dynamic program tracing, and a computer-computer
(A$\leftrightarrow$B) conversational scheme. Compared with the software available for
minicomputers at the time, most of these mechanisms were state of the art.

\begin{figure}[ht]
\begin{center}
\includegraphics[scale=.81]{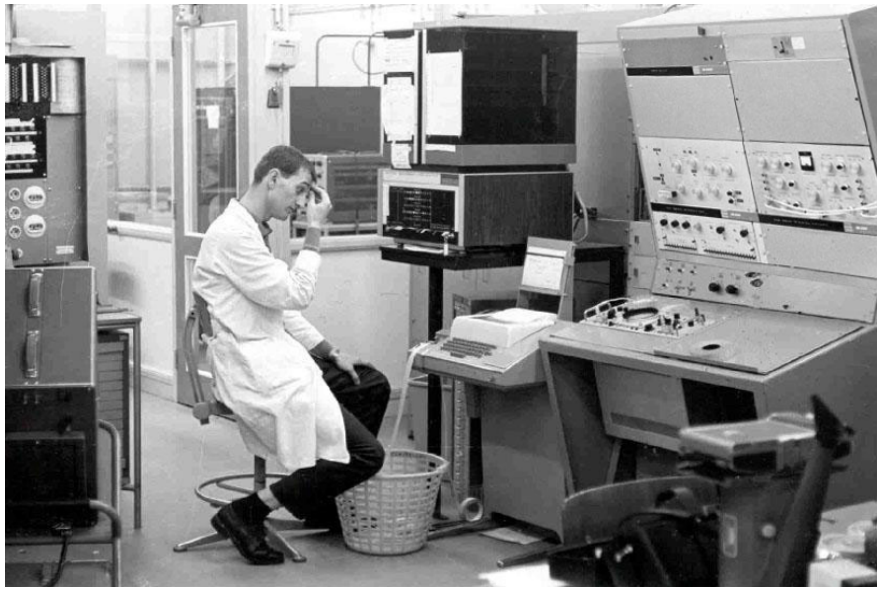}
\end{center} 
\caption{The technician Theo Bijvoets at work with PDP8-A. 
  At right the connected display station. 
The cathode ray tube with light pen is integrated into the desk}
\label{fig:dantzig6}
\end{figure}

In our network, PDP8-A took care of reading the events from BOL, while including
information from the physicist on duty, putting the data in the BOL-format, writing
them on tape, and/or sending them to the X8 and/or to the autonomous display
(Fig.~\ref{fig:dantzig6}) with `light pen'.\footnote{When the light-sensitive pen
touched the oscilloscope screen at an illuminated point, its electric pulses could be
matched with register values determining the position of that point. These values
could then be read out by the PDP8.} Some changes to a M(emory)-unit of a
kicksorter\footnote{A 512 channel Nuclear Data 130 Analyzer, where the least
significant 12 bits were used for data and the 6 most significant bits for logic
functions like interrupting the PDP8, display permission, display intensifying or
light pen flag.} with help from Philips NatLab had made it possible to operate all
previously manually steered functionality of the display by direct computer control
of PDP8-A. With the light pen, we could mark channels independent of the PDP8. An
interface between the M-unit and the display oscilloscope allowed `contour' and
`isometric' 3D-display modes. In the latter case, this was achieved by turning the
viewing angle around. Also, it became possible to make selections on channel address
and/or channel content as well as to adjust the light intensity of each displayed
point based on the memory content. A little later, a storage
oscilloscope\footnote{Tektronix Storage Oscilloscope Type T611.  All interfacing (BOL
detector, tape units and display units) was done by the IKO digital group led by
Evert Kwakkel, which included Pier ten Kate, Jan Kraus, and others.} with hard copy
read-out was interfaced to PDP8-B, thus making it the main display facility.

The PDP8's memory consisting of 32 `pages' of 128 words of 12 bits, allowed direct
addressing within the same page. Memory sharing of asynchronous programs required a
suitable dynamic storage technique.  We divided the core memory into two regions of
pages: the free region and the occupied region. The boundary between them was dynamic
and in fact, the regions could overlap each other page-wise. The administration
contained a list of page-status words. Two functions controlled the memory sharing
administration: `get-page' and `put-page'. Their calls had to be paired off, so that
every job would finish only after giving up all space reserved during runtime.

Whenever a job program was loaded, it obtained the needed pages.  The loader program
itself was also a job program, active in parallel with other programs. Successively
allotted pages were, in general, not contiguous. In order to use the dynamic storage
allocation mechanism for programs, we had to allow programs to be relocatable by page
rather than by program. We restricted references to other pages to jumps and
subroutine calls. This was a limitation we could work with, while the relocation
method could be straightforward and the formats could remain standard.

The format of data streams that have been implemented for the X8, as described in
Section~\ref{sec:dantzig6} in terms of the info-quanta (record, block, train, net,
group, file, repfile), required a similar -- though simplified -- window processing
layer (WSP) for the PDP8s.  For this purpose, a general input-output file handling
system, \footnote{\usereference{Biekman1968}.} tailored to the Monikor system, was
built by our software group.  This system covered all information channels like the
BOL set-up, tape units, the display units, and the X8. Interacting asynchronous
input- and output-processes could `switch' information at the file level between one
input channel and an arbitrary set of output channels. The data stream, which was
divided into the info-quanta, could pass `window programs' as described for the X8.
Specified at the highest protocol level, these could inspect and operate upon the
data between input and output. Different I/O processes could run simultaneously,
while output jobs waiting for input information did not take up processor time.
Synchronization was accomplished using I/O semaphore-functions: the WSP system turned
out to be sufficiently flexible and powerful.

As brought up in Section~\ref{sec:dantzig5}, the X8 tracer had been made in
accordance with our simple method for the PDP8, the `delayed self-interrupt'.  It
consisted of an external interrupt inducing flag (bit), which was cleared at the
entry and set before the exit of a tracer interrupt subroutine, the latter with an
intrinsic delay of about $5 \mu s$.  The delay time was chosen so that only one
instruction of the program under tracing could be executed, at the completion of
which, the flag provided a new program interrupt. While tracing a program, the
interrupt subroutine (possibly adapted to personal needs) could, for example, cause a
stop when the content of a selected register or memory address satisfied a preset
condition. This provided an excellent new test facility for the in-house software
development.

Around 1970, inspired by the design of Wammes and experienced in building the PDP8
Monikor and WSP-system, Anton Mars, Jan Visschers and Ruud van Wijk developed a
genuine time-sharing operating system AIDA\footnote{AIDA = Algemeen Interactief
Data-analyse systeem.} for  a PDP8. I think, this was unprecedented for such a small
machine.  It could serve up to 12 users with time and memory sharing through
alphanumerical terminals. Since the PDP8 was interfaced to the X8, these terminals
also could serve programs running under Wammes, thus extending the number of
simultaneous X8-users.  Relocatable user programs, as described for Monikor, could be
controlled and run independently and simultaneously on the PDP8. All this could be
realized thanks to a hardware extension to the PDP8-processor, a memory protection
unit, designed by the IKO digital group with specifications from the software group.
This unit functionally blocked I/O- and halt-instructions as well as all memory
reference instructions, writing outside certain dynamically specifiable areas of core
memory. Upon occurrence of such a blocking, an interrupt was generated, enabling the
operating system to take appropriate action. AIDA brought considerable satisfaction
for the programmers as much as for the users. It has been used until the DEC-system
10 came into view.

Practically all of the PDP8 software mentioned above was reported at meetings of the
Digital Equipment User Community (DECUS) Europe in
1968\footnote{\usereference{Dantzig1968a}} and in
1972.\footnote{\usereference{Mars1972}; \usereference{Kate1972}.} \section{Final
considerations and conclusions} \label{sec:dantzig8} Over three decades the Philips
Company and IKO had an invaluable collaboration, that included construction and
support of the cyclotron, support on building and maintaining an electron accelerator
as well as the development and production of nuclear
detectors.\footnote{\usereference{Waalwijk1980}. All articles in this issue.}

The Mathematical Center, meeting the growing computational requirements at IKO in the
first half of the 1960s, was essential. The access to the X1 allowed IKO physicists
and technicians to complete practically all of their computer work, amounting to many
hundreds of hours each year.

The continuation of the MC computer developments within Electrologica led to the
production of a substantial series of X8 machines, one of which found its way to IKO.
To obtain the funding, our institute had a strong hand with the portfolio of its
cyclotron, the 85 MeV linear accelerator, a much more powerful 300 MeV electron
accelerator in the planning stage, and last but not least the immediate needs of the
BOL project.  Equally compelling was that Electrologica, as the first Dutch computer
manufacturer, produced a first-class computer and had the ability and willingness to
provide the necessary IKO special hardware and software at reasonable costs.

The X8 made it possible for us to dive into the then modern programming developments
and, together with Electrologica, to create an avant-garde online network with data
processing software.  Programming in ELAN gave us great satisfaction because of its
structure and power. In our view, this assembly language was very well thought-out!

The IKO software group, which was created with the arrival of the X8, was as
enthusiastic and ambitious as we, the physicists and hardware technicians, were. A
major result was Wammes. Time-sharing became indispensable for all of us in many
respects. It gave us ample opportunity to `get at the machine', not only in cases of
experimental urgency, but also for efficient analysis as well as program development
and online testing. Moreover, it allowed us to learn and employ new techniques, such
as list-processing and formula manipulation, thereby accelerating and improving our
work. Wammes may have been the first fully operational time-sharing system in Europe.
Although only tailored to EL-X8-computers, Wammes was in several respects comparable
to the AT\&T Unix operating system which came into worldwide use a few years later.

There were no significant problems in building and operating the BOL network. This
was quite remarkable, since there were two suppliers, the Electrologica staff
involved on the X8 side, and IKO (the BOL-team), on the PDP8 side. The two groups
made an excellent match!

In the end, what really counted was the data processing and analysis.  Of the main
software packages programmed by Electrologica, the X8 Window system was the most
general and indispensable one. It consisted of a set of subroutines and rules that
could have been programmed in any computer language. The fact that there was no
future for Electrologica, for Wammes and for the Window system is regrettable. On the
other hand, we were early workers on -- in contemporary jargon -- `big data' and
`data mining'. That some of us then already got the opportunity to learn and apply
early ideas of `Artificial Intelligence' (AI),\footnote{An `Artificial Intelligence'
study group was formed by, among others, Bob Wielinga and Theo de Ridder.  Bob
Wielinga became a professor at several universities, well known for his AI research
on the methodology of knowledge-based system design and knowledge acquisition.} is
notable, considering that the three areas are now, half a century later, strongly
intertwined.

\begin{figure}[ht]
\begin{center}
\includegraphics[scale=1.03]{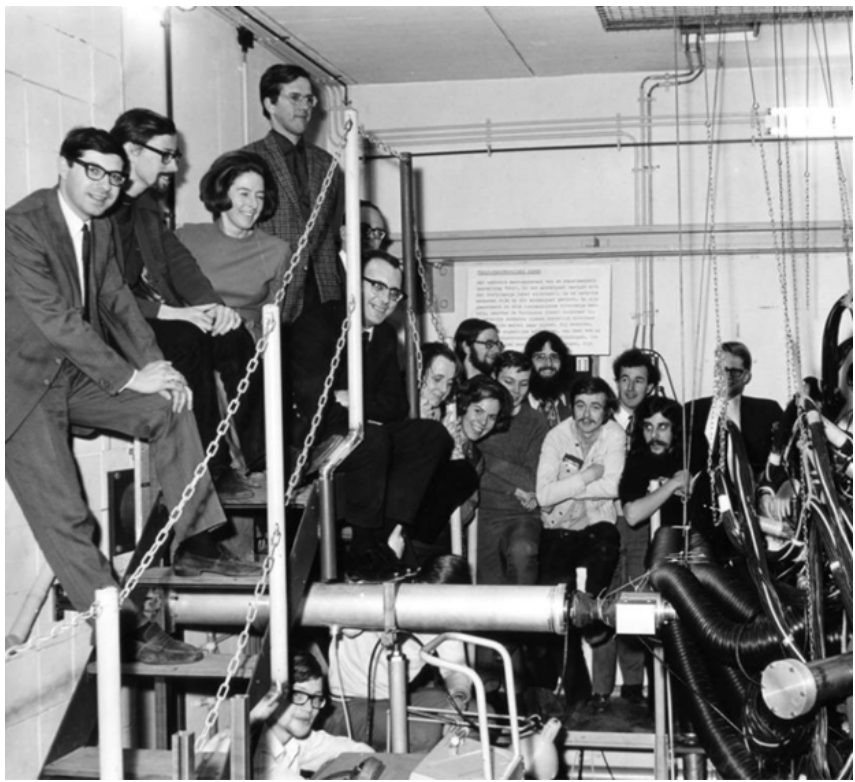}
\end{center}
\caption{Physicists of the BOL-team, 1967: from left to right: Ren\'{e} 
van Dantzig, Ton Ypenberg, Henriette v/d Pijl (secretary), Wil Carton 
(X8-manager), Wim Hermans, Leo Koerts, Theo de Ridder, Jan Lindhout, 
Jos\'{e} Magendans, Jan Waal, Gerard Blommestijn, JEP de Bie, Ton Sonnemans, 
Bob Wielinga, Karel Mulder, Jona Oberski (not visible), Jan Joosten 
(at the bottom)}
\label{fig:dantzig8}
\end{figure}

After having learned about the many positive things of BOL, one might wonder if
everything was really so exciting and successful. Of course, there were setbacks and
challenges.  Parts of the project took much longer than foreseen or were not realized
at all. The cyclotron beam was less intense and could not be focused on targets as
well as had been anticipated. The complexity and size of the electronics required
very time-consuming maintenance. We always had a number of detection units missing,
because they were being repaired, or subtle errors turned up when mounted units were
checked. However, we could live with this, thanks to a lot of redundancy in the
measurement space and -- importantly -- by comparing the experimental data with
adequately simulated theoretical data. In spite of the many problems, the whole
concept of BOL worked; the results, often multidimensional, set a new standard and
were fascinating! To report on our multidimensional results in articles and at
conferences in a way comprehensible to colleagues not familiar with BOL, made that we
had to develop special ways of presentation.\footnote{On this we got advise from
distinguished colleagues, in particular prof.~Ivo \v{S}laus, from Zagreb University.}

\begin{figure}[ht]
\begin{center}
%\sidecaption[c]
\begin{minipage}{0.47\linewidth}
{\includegraphics[scale=0.9]{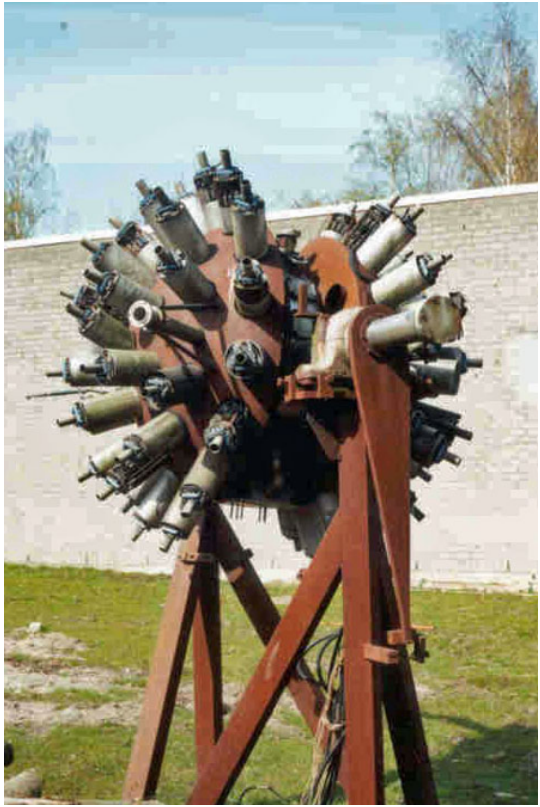}}\end{minipage}
\hspace*{0.5cm}
\begin{minipage}{0.47\linewidth}{\includegraphics[scale=.9]
{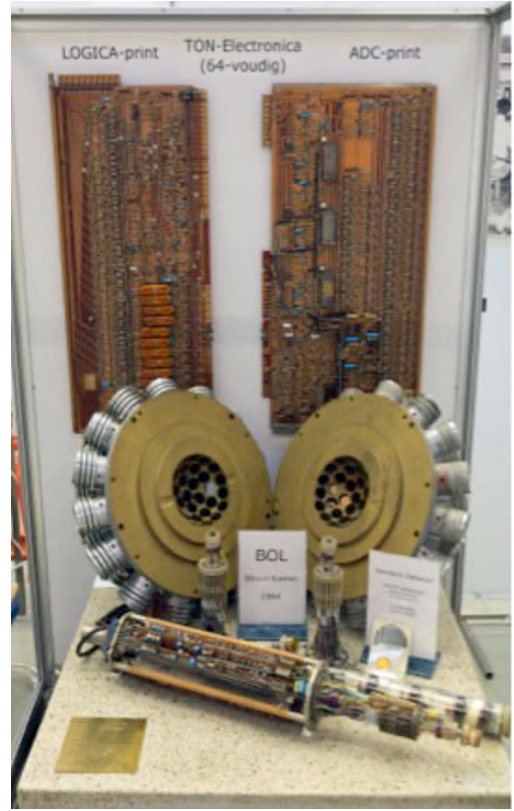}}\end{minipage}
\caption{[At left]: The BOL `monument' on the Science Park  
campus after 25 years, ready for dismantling.
[At right]: Recovered essential parts of BOL on show for 14 years at Nikhef 
  with an explanatory presentation. Visible are the two inner halves of the 
  heart of the BOL-system together with the detectors and electronics for one 
  of the 64 detection channels. In 2016, these parts were included in the 
  collection of the National Museum Boerhaave in Leiden.}
\label{fig:dantzig9}
\end{center}
\end{figure}

We, physicists of the BOL team (Fig.~\ref{fig:dantzig8}), had a mood and attitude,
that somehow belonged to the 1960s: everything was possible or at least should have
been possible! That meant freedom and cooperation in daring experimentation in
physics, electronics and programming.  It also meant being smart, working hard and
with dedication, and doing night shifts at the cyclotron or on the computer after a
full day of work. We were eager to work at the forefront of science and technology,
we read a lot of professional literature and we had our own colloquium and working
group. There was a lot of liberty in choosing the subjects of learning and of
experimentation, even if the connection to the project was uncertain. We became
accustomed to `out of the box' thinking. Occasionally, some of us spent evenings and
nights together and with friends, listening to contemporary pop music, while fleeting
romances could arise with wine and cannabis as catalysts. In summary, I believe that
the multifaceted open-minded spirit of our team contributed a great deal to the BOL
project's success, on top of the acquired expertise, and, to be honest, quite a bit
of good luck, too.

It was in 1971 that we, the BOL `triad', Karel Mulder, Jona Oberski and I, in a full
afternoon marathon ceremony at the University of Amsterdam, successfully defended,
one after the other, our Ph.D theses with BOL results. Our experimental work
continued until 1973, when the measurements were stopped. A total of nearly 2,000
data tapes remained. The data analysis was continued mainly by three graduate
students and several undergraduates. The following year, the X8 was replaced by a
model 10 computer from Digital Equipment Corporation
(DEC).\footnote{\usereference{Dantzig1973}; \usereference{Wielinga1973}.} The main
functionality of the X8 software was reprogrammed in Fortran by staff from DEC.
Analysis and preparations of publications continued officially until 1977. Our last
publication appeared in 1982.\footnote{\useabbrreference{Blommestijn1981}.} The BOL
project might be summarized as a difficult and successful undertaking, with most of
its significance at the forefront of few-body nuclear physics. It led to twenty
publications in international scientific journals, three patents, six PhD-theses, as
well as many master's theses, conference articles, and laboratory reports.

As a whole, BOL with its full infrastructure can be viewed as an early precursor of
modern particle physics set-ups, with its 4$\pi$-type detection `all around' the
collision region, with its silicon detectors for `accurate' impact position
measurements, with its `large' number of electronic channels, with a computer network
online to the experiment and sophisticated software. All of these qualities are, in
extremely scaled-up and improved fashion, characteristics of current high-energy
particle physics experiments.

In 1977, the BOL detector was placed as a `monument' on our campus and stayed there
for 25 years (Fig.~\ref{fig:dantzig9}). Shortly after the turn of the century, it was
dismantled and the heart of the system with examples of its main elements were
exhibited at Nikhef in a showcase with a continuously running presentation until
2016. In that year, the BOL remnants and our corresponding archive were transferred
to the Dutch Science Rijksmuseum Boerhaave in Leiden. That same Museum decided in
2018 to also include an X8, which  had been kept elsewhere,\footnote{This X8 had been
operational at the University of Kiel, Germany.} in the collection. It feels good
that, now, a half-century later, our BOL and an X8 have been saved for posterity,
together.

\paragraph{Acknowledgement}
Much of the chronological framework of this article could be retrieved thanks to IKO
and FOM annual reports.  The black-and-white photographs from the BOL-period, shown
here, were taken by Hans Arnold. I am very much indebted to my former colleagues
from the X8-period and BOL period, who shared with me their recollections and
provided useful comments on versions of this article, in particular JEP de Bie,
Gerard Blommestein, Theo Bijvoets, Pieter van Engen, Anton Mars, Jona Oberski, Henk
Peek, Theo de Ridder, Ton Sonnemans and Jan Visschers. I am especially grateful to
David van Dantzig and Zachary Tobin for textual improvements.  
I feel honored to have been invited by Gerard Alberts, editor of the Springer 
series History of Computing, to compose this article.\\ 
\\This review is dedicated to the memory of my dear brother Émile C.~van Dantzig, 
who started his career in information technology with the X8 at the Mathematical 
Center in 1968 and moved to IKO in 1972, where he enthusiastically and skillfully 
managed the X8 and the succeeding DEC system-10 until 1979.
%\end{acknowledgement}

%\include{chapter}
%\input{references}

%%%%%%%%%%%%%%%%%%%%%%%%%%%%%%%%%%%%%%%%%%%%%%%%%%%%%%%%%%%%%%%%%%%%%%

\end{document}

%% file: definitions_of_references.tex
% !TEX root =  dantzig_article.tex 
\newcommand{\definereference}[3]{
   \expandafter\newcommand\csname Reference#1\endcsname{#2}
   \expandafter\newcommand\csname AbbreviatedReference#1r\endcsname{#3}
}

\newcommand{\usereference}[1]{\csname Reference#1\endcsname}
\newcommand{\useabbrreference}[1]{\csname AbbreviatedReference#1r\endcsname}

\newcommand{\finalreference}[1]{
\bibitem{#1}
\usereference{#1}.
}

%%%%%%%%%%%%%%%%%%%%%%%%%%%%%%%%%%%%%%%%%%%%%%%%%%%%%%%%%%%%%%%%%%%%%%%%%%%%%%%%%%%%%%%%%%%%%%%%%%%%%%%%%%%
\definereference{Biekman1968}
{Biekman, W.C.M. and Anton J.~Mars (1968) PDP-8 IN-UIT-systeem. IKO informal report. 
Archived at the National Museum Boerhaave, Leiden}
{Biekman (1968)}

%%%%%%%%%%%%%%%%%%%%%%%%%%%%%%%%%%%%%%%%%%%%%%%%%%%%%%%%%%%%%%%%%%%%%%%%%%%%%%%%%%%%%%%%%%%%%%%%%%%%%%%%%%%
\definereference{Blommestijn1981}
{Blommestijn, G.J.F., R.~van Dantzig, Y.~Haitsma and R.B.M.~Mooy (1981)
The reaction $d(p, \mathit{pp})n$ measured with BOL at $E_p = 50$ MeV. \textit{Nucl. Phys. A} 365:202-228}
{Blommestijn, e.a. (1981)}

%%%%%%%%%%%%%%%%%%%%%%%%%%%%%%%%%%%%%%%%%%%%%%%%%%%%%%%%%%%%%%%%%%%%%%%%%%%%%%%%%%%%%%%%%%%%%%%%%%%%%%%%%%%
\definereference{Dantzig1967a}
{Dantzig, Ren\'{e} van, et al. (1967a) 
Kwantisering van infostromen: draft. Archived  at the National Museum Boerhaave, Leiden}
{Dantzig, van (1967a)}

%%%%%%%%%%%%%%%%%%%%%%%%%%%%%%%%%%%%%%%%%%%%%%%%%%%%%%%%%%%%%%%%%%%%%%%%%%%%%%%%%%%%%%%%%%%%%%%%%%%%%%%%%%%
\definereference{Dantzig1967b}
{Dantzig, Ren\'{e} van, et al. (1967b) 
Window-processing for on-and off-line data-handling: draft specifications. Archived at the National Museum Boerhaave, Leiden}
{Dantzig, van (1967b)}

%%%%%%%%%%%%%%%%%%%%%%%%%%%%%%%%%%%%%%%%%%%%%%%%%%%%%%%%%%%%%%%%%%%%%%%%%%%%%%%%%%%%%%%%%%%%%%%%%%%%%%%%%%%
\definereference{Dantzig1968a}
{Dantzig, Ren\'{e}~van, et al. (1968a) 
Recent hardware and software developments for the PDP-8. 
Proceedings of the 4th European DECUS Seminar (1968). Archived at the National Museum Boerhaave, Leiden and at Nikhef, Amsterdam.
See also \url{https://ub.fnwi.uva.nl/computermuseum/pdfs/RecentHardwSoftwPDP8.pdf}}
{Dantzig, van (1968a)}

%%%%%%%%%%%%%%%%%%%%%%%%%%%%%%%%%%%%%%%%%%%%%%%%%%%%%%%%%%%%%%%%%%%%%%%%%%%%%%%%%%%%%%%%%%%%%%%%%%%%%%%%%%%
\definereference{Dantzig1968b}
{Dantzig, Ren\'{e}~van, and Stef~Toenbreker (1968b) 
Vensterprogrammering. Archived at the National Museum Boerhaave, Leiden}
{Dantzig, van (1968b)}

%%%%%%%%%%%%%%%%%%%%%%%%%%%%%%%%%%%%%%%%%%%%%%%%%%%%%%%%%%%%%%%%%%%%%%%%%%%%%%%%%%%%%%%%%%%%%%%%%%%%%%%%%%%
\definereference{Dantzig1971a}
{Dantzig, Ren\'{e}~van, et al. (1971a)  
Data acquisition with the BOL nuclear detection system.  \textit{Nucl. Instr. Meth.} 92:199-203}
{Dantzig, van (1971a)}

%%%%%%%%%%%%%%%%%%%%%%%%%%%%%%%%%%%%%%%%%%%%%%%%%%%%%%%%%%%%%%%%%%%%%%%%%%%%%%%%%%%%%%%%%%%%%%%%%%%%%%%%%%%
\definereference{Dantzig1971b}
{Dantzig, Ren\'{e}~van, et al. (1971b) Analysis of multidimensional nuclear data (BOL). \textit{Nucl. Instr. Meth.} 92:205-213}
{Dantzig, van (1971b)}

%%%%%%%%%%%%%%%%%%%%%%%%%%%%%%%%%%%%%%%%%%%%%%%%%%%%%%%%%%%%%%%%%%%%%%%%%%%%%%%%%%%%%%%%%%%%%%%%%%%%%%%%%%%
\definereference{Dantzig1973}
{Dantzig, Ren\'{e}~van, et. al. (1973) 
Voorstel tot vervanging van de EL-X8 computer configuratie van het IKO. 
Archived at the National Museum Boerhaave, Leiden. \url{https://ub.fnwi.uva.nl/computermuseum/pdfs/X8vervanging.pdf}}
{Dantzig, van (1973)}

%%%%%%%%%%%%%%%%%%%%%%%%%%%%%%%%%%%%%%%%%%%%%%%%%%%%%%%%%%%%%%%%%%%%%%%%%%%%%%%%%%%%%%%%%%%%%%%%%%%%%%%%%%%
\definereference{Dantzig1980}
{Dantzig, Ren\'{e}~van (1980) BOL. \textit{Philips Technisch Tijdschrift}. 39/10:286-291 (Dutch, English and French edition)}
{Dantzig, van (1980)}

%%%%%%%%%%%%%%%%%%%%%%%%%%%%%%%%%%%%%%%%%%%%%%%%%%%%%%%%%%%%%%%%%%%%%%%%%%%%%%%%%%%%%%%%%%%%%%%%%%%%%%%%%%%
\definereference{Jonkers1964a}
{Jonkers, H.L., W.T.H.~van Oers, H.R.E.~Tjin A~Djie (1964a)
Discussie rond huidige en toekomstige behoefte aan rekencapaciteit van het IKO. 
IKO internal report Interiko 64. Archived at the National Museum Boerhaave, Leiden and at Nikhef, Amsterdam}
{Jonkers (1964a)}

%%%%%%%%%%%%%%%%%%%%%%%%%%%%%%%%%%%%%%%%%%%%%%%%%%%%%%%%%%%%%%%%%%%%%%%%%%%%%%%%%%%%%%%%%%%%%%%%%%%%%%%%%%%
\definereference{Dantzig2017}
{Dantzig, Ren\'{e}~van (2017) 
Project BOL (1964--1977), Historical overview. \url{https://www.nikhef.nl/history/bol-EN}}
{Dantzig, van (2017)}

%%%%%%%%%%%%%%%%%%%%%%%%%%%%%%%%%%%%%%%%%%%%%%%%%%%%%%%%%%%%%%%%%%%%%%%%%%%%%%%%%%%%%%%%%%%%%%%%%%%%%%%%%%%
\definereference{Dijkstra1968}
{Dijkstra, Edsger~W. (1968) 
The structure of the THE-multiprogramming system. 
Commun. ACM 11(5):341-346.  \url{https://www.cs.utexas.edu/users/EWD/ewd01xx/EWD196.PDF}}
{Dijkstra (1968)}

%%%%%%%%%%%%%%%%%%%%%%%%%%%%%%%%%%%%%%%%%%%%%%%%%%%%%%%%%%%%%%%%%%%%%%%%%%%%%%%%%%%%%%%%%%%%%%%%%%%%%%%%%%%
\definereference{Engen1969}
{Engen, Pieter G.~van, and Rolf Meesters (1969) 
WAMMES, een time-sharing systeem voor de EL-X8: gids voor gebruikers. 
Archived at the National Museum Boerhaave, Leiden. \url{https://ub.fnwi.uva.nl/computermuseum/pdfs/WAMMESgebruikersgids.pdf}}
{Engen (1969)}

%%%%%%%%%%%%%%%%%%%%%%%%%%%%%%%%%%%%%%%%%%%%%%%%%%%%%%%%%%%%%%%%%%%%%%%%%%%%%%%%%%%%%%%%%%%%%%%%%%%%%%%%%%%
\definereference{Jonkers1964b}
{Jonkers, H.L. (1964b) 
A comparison EL-X8 and CDC-3200 for on-line use. 
IKO internal report Interiko 64/5. Archived at the National Museum Boerhaave, 
Leiden and at Nikhef, Amsterdam. \url{https://ub.fnwi.uva.nl/computermuseum/pdfs/X8vsCDC3200.pdf}}
{Jonkers (1964b)}

%%%%%%%%%%%%%%%%%%%%%%%%%%%%%%%%%%%%%%%%%%%%%%%%%%%%%%%%%%%%%%%%%%%%%%%%%%%%%%%%%%%%%%%%%%%%%%%%%%%%%%%%%%%
\definereference{Jonkers1964c}
{Jonkers, H.L. (1964c) 
Computer requirements of the Institute for Nuclear Physics Research. 
IKO informal report. Archived at the National Museum Boerhaave, Leiden and at Nikhef, Amsterdam}
{Jonkers (1964c)}

%%%%%%%%%%%%%%%%%%%%%%%%%%%%%%%%%%%%%%%%%%%%%%%%%%%%%%%%%%%%%%%%%%%%%%%%%%%%%%%%%%%%%%%%%%%%%%%%%%%%%%%%%%%
\definereference{Heijne2003}
{Heijne, Erik H.M. (2003) Semiconductor detectors in the low countries. 
\textit{Nucl. Instr. Meth. Phys. A} 309:1-16}
{Heijne (2003)}

%%%%%%%%%%%%%%%%%%%%%%%%%%%%%%%%%%%%%%%%%%%%%%%%%%%%%%%%%%%%%%%%%%%%%%%%%%%%%%%%%%%%%%%%%%%%%%%%%%%%%%%%%%%
\definereference{Heyn1952}
{Heyn, F.A., and J.J.~Burgerjon (1952)  
Het synchro-cyclotron te Amsterdam. \textit{Philips Technisch Tijdschrift} 14:291-307} 
{Heyn (1952)}

%%%%%%%%%%%%%%%%%%%%%%%%%%%%%%%%%%%%%%%%%%%%%%%%%%%%%%%%%%%%%%%%%%%%%%%%%%%%%%%%%%%%%%%%%%%%%%%%%%%%%%%%%%%
\definereference{Kate1972}
{Kate, P.U. ten, and E.~Kwakkel (1972) 
AIDA - An operating system for the PDP-8. II. Hardware. 
\textit{Proceedings 8th DECUS Europe Seminar 1972}. Archived at Nikhef, Amsterdam}
{Kate, ten (1972)}

%%%%%%%%%%%%%%%%%%%%%%%%%%%%%%%%%%%%%%%%%%%%%%%%%%%%%%%%%%%%%%%%%%%%%%%%%%%%%%%%%%%%%%%%%%%%%%%%%%%%%%%%%%%
\definereference{Koerts1971}
{Koerts, L.A.Ch., K.~Mulder, J.E.J.~Oberski and R.~van Dantzig (1971) 
The BOL nuclear research project. \textit{Nucl. Instr. Meth.} 92:157-160}
{Koerts (1971)}

%%%%%%%%%%%%%%%%%%%%%%%%%%%%%%%%%%%%%%%%%%%%%%%%%%%%%%%%%%%%%%%%%%%%%%%%%%%%%%%%%%%%%%%%%%%%%%%%%%%%%%%%%%%
\definereference{Luijckx1980}
{Luijckx, Guy (1980) Het cyclotron. \textit{Philips Technisch Tijdschrift} 39(10):274-276}
{Luijckx (1980)}

%%%%%%%%%%%%%%%%%%%%%%%%%%%%%%%%%%%%%%%%%%%%%%%%%%%%%%%%%%%%%%%%%%%%%%%%%%%%%%%%%%%%%%%%%%%%%%%%%%%%%%%%%%%
% \url{https://en.wikipedia.org/wiki/University\_of\_Amsterdam}
\definereference{Mars1972}
{Mars, Anton J., J.L.~Visschers and R.F. van Wijk (1972) 
AIDA -- An operating system for the PDP-8. I. Software. 
\textit{Proceedings of the 8th DECUS Europe Seminar 1972}. Archived at Nikhef, Amsterdam}
{Mars (1972)}

%%%%%%%%%%%%%%%%%%%%%%%%%%%%%%%%%%%%%%%%%%%%%%%%%%%%%%%%%%%%%%%%%%%%%%%%%%%%%%%%%%%%%%%%%%%%%%%%%%%%%%%%%%%
\definereference{Oberski1965}
{Oberski, Jona E.J., R.~van Dantzig, K.~Mulder, L.A.Ch.~Koerts, and F.~van Hall (1965) 
IKOB, koppeling van PDP8 en X8. IKO Internal report. Archived at Nikhef, Amsterdam}
{Oberski (1965)}

%%%%%%%%%%%%%%%%%%%%%%%%%%%%%%%%%%%%%%%%%%%%%%%%%%%%%%%%%%%%%%%%%%%%%%%%%%%%%%%%%%%%%%%%%%%%%%%%%%%%%%%%%%%
\definereference{Oberski1967}
{Oberski, Jona E.J, L.A.Ch.~Koerts, R.~van Dantzig and K.~Mulder (1967) 
Toelichting op de computers in BOL. 
IKO internal report Interiko 67/9.  
Archived at the National Museum Boerhaave, Leiden and at Nikhef, Amsterdam. \url{https://ub.fnwi.uva.nl/computermuseum/pdfs/computersBOL.pdf}}
{Oberski (1967)}

%%%%%%%%%%%%%%%%%%%%%%%%%%%%%%%%%%%%%%%%%%%%%%%%%%%%%%%%%%%%%%%%%%%%%%%%%%%%%%%%%%%%%%%%%%%%%%%%%%%%%%%%%%%
\definereference{Oberski1971a}
{Oberski, Jona E.J., et al. (1971a) Electronics of the BOL System. \textit{Nucl. Instr. Meth.} 92:177-187}
{Oberski (1971a)}

%%%%%%%%%%%%%%%%%%%%%%%%%%%%%%%%%%%%%%%%%%%%%%%%%%%%%%%%%%%%%%%%%%%%%%%%%%%%%%%%%%%%%%%%%%%%%%%%%%%%%%%%%%%
\definereference{Oberski1971b}
{Oberski, Jona E.J. et al. (1971b) The BOL Computer Hardware Configuration. \textit{Nucl. Instr. Meth.} 92:189-191}
{Oberski (1971b)}

%%%%%%%%%%%%%%%%%%%%%%%%%%%%%%%%%%%%%%%%%%%%%%%%%%%%%%%%%%%%%%%%%%%%%%%%%%%%%%%%%%%%%%%%%%%%%%%%%%%%%%%%%%%
\definereference{Ridder1972} 
{Theo F.~de Ridder (1972) LISI, een LISP-SIMPLEX systeem met display faciliteiten, Informal IKO report. Archived at the National Museum Boerhaave, Leiden}
{Ridder (1972)}

%%%%%%%%%%%%%%%%%%%%%%%%%%%%%%%%%%%%%%%%%%%%%%%%%%%%%%%%%%%%%%%%%%%%%%%%%%%%%%%%%%%%%%%%%%%%%%%%%%%%%%%%%%%
\definereference{Toenbreker1965}
{Toenbreker, S.J.A.M., P.G.~van Engen, R.~van Dantzig, L.A.Ch.~Koerts, K.~Mulder, and J.E.J.~Oberski (1965) 
De toepassing van een EL X-8 op het IKO. 
IKO internal report Interiko 65/9. 
Archived at the National Museum Boerhaave, Leiden and at Nikhef, Amsterdam. \url{https://ub.fnwi.uva.nl/computermuseum/pdfs/X8opIKO.pdf}}
{Toenbreker (1965)}

%%%%%%%%%%%%%%%%%%%%%%%%%%%%%%%%%%%%%%%%%%%%%%%%%%%%%%%%%%%%%%%%%%%%%%%%%%%%%%%%%%%%%%%%%%%%%%%%%%%%%%%%%%%
\definereference{Wapstra1964}
{Wapstra, A.H., H.L.~Jonkers, and L.A.Ch.~Koerts (1964) 
Internal note `Rekenmachine'. \url{https://ub.fnwi.uva.nl/computermuseum/pdfs/Wapstra.pdf}}
{Wapstra, Jonkers and Koerts (1964), `Rekenmachine'}

%%%%%%%%%%%%%%%%%%%%%%%%%%%%%%%%%%%%%%%%%%%%%%%%%%%%%%%%%%%%%%%%%%%%%%%%%%%%%%%%%%%%%%%%%%%%%%%%%%%%%%%%%%%
\definereference{Wapstra1965}
{Wapstra, A.H., H.L.~Jonkers, et al. (1965) 
Voorstel tot aanschaffing van een electronische rekenautomaat ten behoeve van het fysisch werk op het Instituut voor Kernfysisch Onderzoek. 
Internal report IKO-2168. \url{https://ub.fnwi.uva.nl/computermuseum/pdfs/X8aanschafvoorstel.pdf}}
{Wapstra (1965)}

%%%%%%%%%%%%%%%%%%%%%%%%%%%%%%%%%%%%%%%%%%%%%%%%%%%%%%%%%%%%%%%%%%%%%%%%%%%%%%%%%%%%%%%%%%%%%%%%%%%%%%%%%%%
\definereference{Wielinga1973}
{Wielinga, B.J., J.E.J.~Oberski, R.~van Dantzig et al. (1973) 
Toelichting op de aanvraag voor vervanging van de rekenmachine van het Instituut voor Kernphysisch Onderzoek (IKO) te Amsterdam. 
IKO informal report. 
Archived at the National Museum Boerhaave, Leiden. \url{https://ub.fnwi.uva.nl/computermuseum/pdfs/vervanging1973.pdf}}
{Wielinga (1973)}